\documentclass[prd,preprint,superscriptaddress,amsmath,amssymb,nofootinbib]{revtex4}
\usepackage{graphicx}
\usepackage{dcolumn}
\usepackage{bm}
\usepackage{amssymb}
\usepackage{amsmath}
\usepackage{epsfig}    
\usepackage{color}
\usepackage{slashed}
\usepackage{hhline}

\def\be{\begin{equation}}
\def\ee{\end{equation}}
\newcommand{\bea}{\begin{eqnarray}}
\newcommand{\eea}{\end{eqnarray}}

\begin{document}

\title{Neutrino observables in gauged $U(1)_{L_\alpha - L_\beta}$ models with two Higgs doublet and one singlet scalars}

\author{Yuanchao Lou}
\email{louyuanchao@stu.scu.edu.cn}
\affiliation{College of Physics, Sichuan University, Chengdu 610065, China}

\author{Takaaki Nomura}
\email{nomura@scu.edu.cn}
\affiliation{College of Physics, Sichuan University, Chengdu 610065, China}

\date{\today}

\begin{abstract}
We discuss neutrino sector in models with two Higgs doublet and one singlet scalar fields under local $U(1)_{L_\alpha- L_\beta}$ symmetry.
A neutrino mass matrix is formulated for these models where the matrix is generated via type-I seesaw mechanism introducing right-handed neutrinos.
The neutrino mass matrix has more degrees of freedom compared to minimal scenarios which have only one new scalar field, but its structure is still restricted by the symmetry.
Then it is find that sum of neutrino mass can be lower than minimal scenarios and it is easier to satisfy observed constraints.
In addition, we can fit neutrino data for $U(1)_{L_e - L_{\mu(\tau)}}$ cases which are disfavored in minimal models. 
Furthermore, some correlations among sum of neutrino mass and CP violating phases are still found although we have more free parameters. 
 \end{abstract}
\maketitle

\section{Introduction}

The nature of neutrino such as non-zero masses and mixings is one of the biggest mystery in particle physics.
We need to extend the standard model (SM) to generate neutrino masses where an attractive way is the type-I seesaw mechanism introducing right-handed neutrinos~\cite{Minkowski:1977sc,Yanagida:1979as,Yanagida:1979gs, Gell-Mann:1979vob,Mohapatra:1979ia, Schechter:1980gr} as a simple possibility.
New physics sector would have more rich structure containing new particle contents in addition to right-handed neutrinos and/or new (gauge) symmetries.
It is thus important to explore compatibility among neutrino mass/mixing and new physics, especially a flavor dependent symmetry.

An introduction of new $U(1)$ gauge symmetry provides simple extensions of the SM.
Among many possibilities local $U(1)_{L_\alpha - L_\beta}$ symmetries are interesting where they are anomaly free and the SM leptons have flavor dependent charge $L_\alpha - L_\beta (\alpha,\beta = \{e,\mu, \tau\})$.
Introduction of such a symmetry restricts the structure of Yukawa interactions and Majorana mass of right-handed neutrinos that are related to neutrino mass structure.
Actually compatibility among the $U(1)_{L_\alpha - L_\beta}$ symmetry and neutrino masses/mixings has been explored for type-I seesaw scenario~\cite{Asai:2018ocx, Asai:2017ryy} where one scalar field, $SU(2)_L$ singlet or doublet with non-zero $U(1)_{L_\alpha - L_\beta}$ charge, is introduced  to develop a vacuum expectation value (VEVs) contributing to neutrino mass via Yukawa interaction as minimal choices~\footnote{One can find other approaches, especially $U(1)_{L_\mu - L_\tau}$ case, discussing constrained neutrino mass matrix in other seesaw models including radiative mass generation~\cite{Baek:2015mna, Lee:2017ekw, Chen:2017gvf, Nomura:2018vfz,Nomura:2018cle, Araki:2019rmw, Asai:2019ciz,Bauer:2020itv, Nomura:2023vmh}. }. 
In this case we have clear relations among neutrino observables and models are restricted by the current neutrino data~\cite{Esteban:2020cvm}.
It is then found that only $U(1)_{L_\mu - L_\tau}$ case with one singlet scalar having charge $\pm 1$ 
can accommodate neutrino data when we consider constraints on sum of neutrino mass $\sum m_\nu \lesssim 0.12$ eV from CMB data by Planck~\cite{Planck:2018vyg} under the standard $\Lambda$CDM cosmological model.
Remarkably more stringent constraint is obtained as $\sum m_\nu < 0.07$ eV (95\%) if we include recent baryon acoustic oscillation (BAO) analysis by Dark Energy Spectroscopic Instrument (DESI) data~\cite{DESI:2024mwx}; even stronger bounds are estimated in refs.~\cite{Craig:2024tky, Wang:2024hen}. 
Then $U(1)_{L_\alpha - L_\beta}$ models are excluded when we only add one new scalar field and impose the constraint on the sum; it is marginal even if we consider $\sum m_\nu > 0.059$ eV prior leading bit looser constraint $\sum m_\nu < 0.113$.
These analysis indicate we need to extend minimal $U(1)_{L_\alpha - L_\beta}$ models or modify standard $\Lambda$CDM model.
Thus it is worth considering a next minimal case including both $SU(2)_L$ singlet and doublet scalar fields~\footnote{Such a hybrid type is discussed in ref.~\cite{Asai:2024pzx} to relax constraints regarding $Z'$ boson. Also similar hybrid models are discussed in refs.~\cite{Heeck:2011wj, Heeck:2014qea, Crivellin:2015lwa, Ardu:2022zom} for lepton flavor violation physics.} and investigate if we can still get some predictions in neutrino sector satisfying 
the constraint on neutrino mass at the same time.

In this paper, we discuss models with a local $U(1)_{L_\alpha - L_\beta}$ symmetry introducing right-handed neutrinos, and $SU(2)_L$ doublet and singlet scalars with non-zero charges under the 
new gauge symmetry. 
The models are characterized by the gauge symmetry and possible choices of charges for doublet scalar field.
We then explore neutrino masses and mixings in each model, and search for parameters that can accommodate neutrino data.
For allowed parameter sets we show some predictions such as sum of neutrino mass and CP violating phases.

This paper is organized as follows.
In Sec. II, we introduce models and formulate scalar sector, neutral gauge bosons, charged lepton mass and neutrino mass matrix.
In Sec. III, we show numerical analysis of neutrino mass showing some predictions.
Finally we conclude and discuss in Sec.IV. 
 
\section{Models}

\begin{table}[t!]
\begin{tabular}{|c||c|c|c|c|c|c|c|c|c|}\hline\hline  
& ~$L_{L_i}$~& ~$e_{R_i}$~& ~$\nu_{R_i}$~ & ~$H_1$~&~$H_2$~& ~$\varphi$~ \\\hline\hline 
$SU(2)_L$  & $\bm{2}$ & $\bm{1}$  & $\bm{1}$ & $\bm{2}$ & $\bm{2}$  & $\bm{1}$     \\\hline 
$U(1)_Y$    & $-\frac12$ & $-1$  & $0$ & $1/2$ & $1/2$ & $0$ \\\hline
$U(1)_{L_\alpha - L_\beta}$    & $L_\alpha - L_\beta$ & $L_\alpha - L_\beta$  & $L_\alpha - L_\beta$ & $Q_H$ & $0$ & $1$ \\\hline
\end{tabular}
\caption{Charge assignments of the leptons and scalar fields
under $ SU(2)_L\times U(1)_Y \times U(1)_{L_\alpha - L_\beta}$ where $L_{\alpha,\beta}$ corresponds to lepton number for $\alpha(\beta) = \{e, \mu, \tau \}$ flavor  ($\alpha \neq \beta $) and $Q_H = +1$ or $-1$.   }\label{tab:1}
\end{table}

In this section we introduce models and formulate neutrino mass matrix.
Models are constructed under a framework of two Higgs doublet plus one singlet scalar under $U(1)_{L_\alpha - L_\beta}$ gauge symmetry.
We also introduce three right-handed neutrinos charged under $U(1)_{L_\alpha- L_\beta}$ to realize type-I seesaw mechanism.
In this scenario one Higgs doublet $H_1$ has charge $Q_H = \pm 1$ and singlet scalar has charge $1$ under $U(1)_{L_\alpha - L_\beta}$ where we summarized charge assignment in Table~\ref{tab:1}.
Notice that $Q_H = \pm 1$ is also required to make operator $\varphi H_1^\dagger H_2$(or $\varphi^* H_1^\dagger H_2$) gauge invariant to avoid massless Goldstone boson from Higgs doublets.
In fact these assignments of $U(1)_{L_\alpha -L_\beta}$ charge to scalar fields are only relevant ones to obtain neutrino mass matrix which can fit the neutrino data.

Relevant  Lagrangian for lepton sector is written by 
\begin{align}
L_{\ell} = & \ y_{\ell_1} \bar {L}_L e_R H_1 +   y_{\ell_2} \bar {L}_L e_R H_2 + y_{\nu_1} \bar{L}_L \nu_R \tilde{H}_1 + y_{\nu_2} \bar{L}_L \nu_R \tilde{H}_2 \nonumber \\
&+ \frac12 M_{0} \overline{\nu^c_R} \nu_R  + \frac12 y_M \overline{\nu^c_R} \nu_R \varphi  + \frac12 \tilde{y}_M \overline{\nu^c_R} \nu_R \varphi^*  + h.c. \, ,
\label{eq:Lagrangian}
\end{align}
where $\tilde{H}_i = H_i^* i \sigma_2$ with $\sigma_2$ being the second Pauli matrix and flavor index is omitted.
The structure of Yukawa coupling matrices and bare Majorana mass matrix are constrained by charge assignments under $U(1)_{L_\alpha - L_\beta}$ as we discuss below.
Scalar potential is also given by
\begin{align}
V = & \ m_1^2 H_1^\dagger H_1 + m_2^2 H_2^\dagger H_2 + m_\varphi^2 |\varphi|^2  - (\mu \varphi^{(*)} H_1^\dagger H_2 + h.c.) 
+ \lambda_1 (H_1^\dagger H_1)^2 + \lambda_2 (H_2^\dagger H_2)^2  + \lambda_\varphi |\varphi|^4 \nonumber \\
& + \lambda_3 (H_1^\dagger H_1)(H_2^\dagger H_2) + \lambda_4 (H_1^\dagger H_2)(H_2^\dagger H_1) + \lambda_{H_1 \varphi} (H_1^\dagger H_1) |\varphi|^2 
+  \lambda_{H_2 \varphi} (H_2^\dagger H_2) |\varphi|^2,
\end{align}
where $\varphi$ or $\varphi^*$ is determined by the $U(1)_{L_\alpha - L_\beta}$ charge of $H_1$ for the fourth term.
In this work we consider six models given in Table~\ref{tab:2} that are distinguished by $U(1)$ symmetry and charge $Q_H$ providing us different structure of neutrino mass matrix.

\begin{table}[t]
\begin{tabular}{|c||c|c|c|c|c|c|c|c|c|}\hline\hline  
& ~model (1)~ & ~model (2)~ & ~model (3)~ & ~model (4)~ & ~model (5)~ & ~model (6)~ \\\hline\hline 
$U(1)_{L_i - L_j}$  & $L_e - L_\mu$ & $L_e - L_\mu$  & $L_e - L_\tau$ & $L_e - L_\tau$ & $L_\mu - L_\tau$  & $L_\mu - L_\tau$     \\\hline 
$Q_H$    & $1$ & $-1$  & $1$ & $-1$ & $1$ & $-1$ \\\hline
\end{tabular}
\caption{Models distinguished by extra $U(1)$ symmetry and $H_1$ charge $Q_H$.   }\label{tab:2}
\end{table}

\subsection{Scalar sector}

Here we review scalar sector containing two Higgs doublet and one singlet scalar under extra $U(1)_{L_\alpha - L_\beta}$ gauge symmetry.
The two Higgs doublets and singlet scalar are represented as
\begin{equation}
H_i = \begin{pmatrix} \phi_a^+ \\ \frac{1}{\sqrt{2}} (v_a + h_a + i \eta_a) \end{pmatrix}, \quad \varphi = \frac{1}{\sqrt{2}} (\phi_R + v_\varphi + i \phi_I),
\end{equation}
where $a=1,2$, and $v_a$ and $v_\varphi$ are the VEVs of the corresponding fields.
The VEVs can be obtained from the stationary conditions $(\partial V/\partial h_i)_0 = (\partial V/\partial \phi_R)_0 = 0$ where subscript $"0"$ indicates 
all the component fields are taken to be zero. 
These conditions provide 
\begin{align}
& m_{H_1}^2 v_1 - m_{12}^2 v_2 + \frac{v_1}{2} (2 v_1^2 \lambda_1 + v_2^2 \bar \lambda + v_\varphi^2 \lambda_{H_1 \varphi}) =0 \nonumber \\
& m_{H_2}^2 v_2 - m_{12}^2 v_1 + \frac{v_2}{2} (2 v_2^2 \lambda_2 + v_1^2 \bar \lambda + v_\varphi^2 \lambda_{H_2 \varphi}) =0 \nonumber \\
& m_\varphi^2 v_\varphi - \frac{1}{\sqrt{2}} \mu v_1 v_2  + \lambda_\varphi v_\varphi^3 + \frac{\lambda_{H_1 \varphi}}{2} v_1^2 v_\varphi + \frac{\lambda_{H_2 \varphi}}{2} v_2^2 v_\varphi =0,
\label{eq:VEVcondition}
\end{align}
where $m_{12}^2 \equiv \mu v_\varphi/\sqrt{2} $ and $\bar \lambda = \lambda_3 + \lambda_4$.

The mass eigenstates for charged scalar components are obtained as in a two Higgs doublet model (THDM), 
\begin{equation}
\begin{pmatrix} G^\pm \\ H^\pm \end{pmatrix} = \begin{pmatrix} \cos \beta & - \sin \beta \\ \sin \beta & \cos \beta \end{pmatrix} \begin{pmatrix} \phi_1^\pm \\ \phi_2^\pm \end{pmatrix},
\end{equation}
where $\tan \beta = v_2/v_1$, $G^\pm$ corresponds to Nambu-Goldstone (NG) boson absorbed by $W^\pm$ and $H^\pm$ is physical charged Higgs boson.
The mass of the charged Higgs boson can be written by
\begin{equation}
m^2_{H^\pm} = \frac{m^2_{12}}{\sin \beta \cos \beta} - \frac{v^2}{2} \lambda_4,
\end{equation}
where $v = \sqrt{v_1^2 + v_2^2} \simeq 246$ GeV.

Applying conditions in Eq.~\eqref{eq:VEVcondition}, the mass matrix for CP-odd scalar bosons is obtained as 
\begin{equation}
\mathcal{L} \in \frac12 \begin{pmatrix} \eta_1 \\ \eta_2 \\ \phi_I \end{pmatrix}^T 
\begin{pmatrix} \frac{\mu v_\varphi v_2}{\sqrt{2} v_1} & - \frac{\mu v_\varphi}{\sqrt{2}} & - \frac{\mu v_2}{\sqrt{2}} \\
 - \frac{\mu v_\varphi}{\sqrt{2}} & \frac{\mu v_\varphi v_1}{\sqrt{2} v_2} &  \frac{\mu v_1 v_2}{\sqrt{2} v_\varphi} \\
 - \frac{\mu v_2}{\sqrt{2}} &  \frac{\mu v_1}{\sqrt{2}} &  \frac{\mu v_2}{\sqrt{2}} 
 \end{pmatrix}
 \begin{pmatrix} \eta_1 \\ \eta_2 \\ \phi_I \end{pmatrix}.
\end{equation}
This mass matrix can be diagonalized by rotating the basis as follows~\cite{Nomura:2019wlo};
\begin{equation}
 \begin{pmatrix} \eta_1 \\ \eta_2 \\ \phi_I \end{pmatrix} = 
 \begin{pmatrix}
 \frac{v_1}{v} & - \frac{v_\varphi v_2}{\sqrt{v_1^2 v_2^2 + v_\varphi^2 v^2}} & \frac{v_1}{\sqrt{v_\varphi^2 + v_1^2}} \\
 \frac{v_2}{v} & - \frac{v_\varphi v_1}{\sqrt{v_1^2 v_2^2 + v_\varphi^2 v^2}} & 0 \\
 0 & - \frac{v_1 v_2}{\sqrt{v_1^2 v_2^2 + v_\varphi^2 v^2}} & \frac{v_\varphi}{\sqrt{v_\varphi^2 + v_1^2}}
 \end{pmatrix}
 \begin{pmatrix} G^0_1 \\ A^0 \\ G^0_2 \end{pmatrix},
\end{equation}
where $G_{1,2}^0$ are massless NG boson whose degrees of freedom are absorbed by $Z$ and $Z'$ bosons.
Here we note that $\{\eta_1, \eta_2 \}$ sector becomes THDM like in the limit of $v_\varphi \gg v$.
The mass eigenvalue of physical CP-odd scalar boson $A^0$ is given by
\begin{equation}
m^2_{A^0} =  \frac{m_{12}^2}{\sin \beta \cos \beta} + \frac{1}{\sqrt{2}} \frac{\mu v^2}{v_\varphi}.
\end{equation}
Note that the mass of $A^0$ becomes zero when we take $\mu =0$ since there is spontaneously broken global $U(1)$ symmetry in the potential for the limit.

The CP-even scalar sector has three physical degrees of freedom $\{h_1, h_2, \phi_R \}$ and the mass matrix is written by
\begin{equation}
\mathcal{L} \supset \frac{1}{2}
\left( \begin{array}{c} h_1 \\ h_2 \\ \phi_R \end{array} \right)^T 
\left(
\begin{array}{ccc}
 2 \lambda_1 v_1^2 + \frac{ \mu v_\varphi   v_2}{\sqrt{2} v_1} & \lambda _3 v_1 v_2+\lambda_4 v_1 v_2 - \frac{  \mu  v_\varphi}{\sqrt{2}} & \eta \lambda_{ H_1 \varphi} v_1 - \frac{\mu  v_2}{\sqrt{2}} \\
 \lambda_3 v_1 v_2+\lambda_4 v_1 v_2 - \frac{ \mu  v_\varphi}{\sqrt{2}} & 2 \lambda_2v_2^2 + \frac{ \mu   v_1 v_\varphi}{\sqrt{2} v_2} & v_\varphi \lambda_{ H_2 \varphi}v_2 -\frac{\mu  v_1}{\sqrt{2}}  \\
 v_\varphi  \lambda_{ H_1 \varphi} v_1 - \frac{\mu v_2}{\sqrt{2}} & v_\varphi \lambda_{ H_2 \varphi} v_2 - \frac{\mu v_1}{\sqrt{2}} & 2 v_\varphi^2 \lambda_{\varphi} + \frac{\mu  v_1 v_2}{\sqrt{2} \varphi } \\
\end{array}
\right) 
\left( \begin{array}{c} h_1 \\ h_2 \\ \phi_R \end{array} \right),
\end{equation}
where we imposed conditions in Eq.~\eqref{eq:VEVcondition}.
This mass matrix can be diagonalized by $3 \times 3$ orthogonal matrix providing three physical mass eigenvalues.
Such an orthogonal matrix $R$ with three Euler parameters $\{ \alpha_1, \alpha_2, \alpha_3 \}$ is written by
\begin{equation}
\\ R(\alpha_1,\alpha_2,\alpha_3)=\left(\begin{array}{ccc}
c_{\alpha_1}c_{\alpha_2} & - s_{\alpha_1}c_{\alpha_2} & s_{\alpha_2} \\
- c_{\alpha_1}s_{\alpha_2}s_{\alpha_3} + s_{\alpha_1}c_{\alpha_3} & c_{\alpha_1}c_{\alpha_3}+s_{\alpha_1}s_{\alpha_2}s_{\alpha_3} & c_{\alpha_2}s_{\alpha_3}\\
-c_{\alpha_1}s_{\alpha_2}c_{\alpha_3}-s_{\alpha_1}s_{\alpha_3} & - c_{\alpha_1}s_{\alpha_3} + s_{\alpha_1}s_{\alpha_2}c_{\alpha_3}& c_{\alpha_2}c_{\alpha_3}\\
\end{array}\right),
\end{equation} 
where $c_{\alpha_i} = \cos \alpha_i (s_{\alpha_i} = \sin \alpha_i)$.
Then mass eigenstates are obtained such that
\begin{equation}
\left( \begin{array}{c} h_1 \\ h_2 \\ \phi_R \end{array} \right) = R_{ij} \left( \begin{array}{c} H^0 \\ h^0 \\ \xi^0 \end{array} \right)_j.
\label{Eq:CP-even}
\end{equation}
In this work, we do not discuss more details of the scalar sector since our focus is neutrino sector.
Thus we simply assume parameters in the scalar potential are chosen to satisfy phenomenological constraints regarding Higgs boson physics.

\subsection{Neutral gauge bosons}

\begin{figure}[tb]
 \begin{center}
\includegraphics[width=8cm]{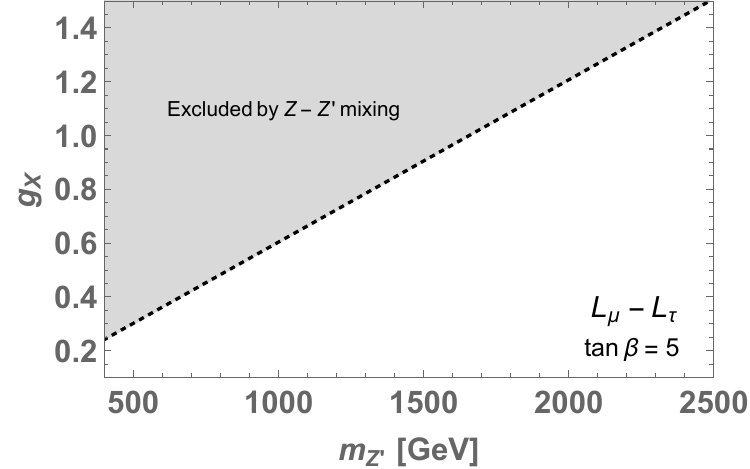}  \
 \caption{The gray region is excluded by EWPT for $U(1)_{L_\mu - L_\tau}$ case with $\tan \beta =5$.}
\label{fig:ZZpMixing}
\end{center}
\end{figure}

Here we focus on neutral gauge boson sector in the model since the charged gauge boson $W^\pm$ is the same as the SM one.
After spontaneous symmetry breaking we obtain mass terms 
\begin{equation}
\mathcal{L}_M = \frac12 m^2_{Z_{\rm SM}} \tilde Z_\mu \tilde Z^\mu + \frac12 \tilde m_{Z'}^2 \tilde Z'_\mu \tilde Z'^\mu + \Delta M^2 \tilde Z_\mu \tilde Z'^\mu,
\end{equation}
where $m^2_{Z_{\rm SM}} = v^2 (g_1^2 + g_2^2)/4$ with $g_{1(2)}$ being $U(1)_Y(SU(2)_L)$ gauge coupling, $\tilde m_{Z'}^2 = g_X^2 (v_\varphi^2+ v_1^2)$ with new gauge coupling $g_X$,
and $\Delta M^2 = g_X \sqrt{g_1^2 + g_2^2}  v_1^2/2$.
Here $\tilde Z_\mu = \cos \theta_W W^3_\mu - \sin \theta_W B_\mu$ where $W^3_\mu$ and $B_\mu$ come from $SU(2)_L$ and $U(1)_Y$ gauge fields as in the SM while 
$\tilde Z'_\mu$ is $U(1)_{L_\alpha - L_\beta}$ gauge field. Notice that we do not consider kinetic mixing among $U(1)$ gauge fields assuming it is negligibly small.
Then we can diagonalize the mass terms by rotating the fields and get mass eigenstates  
\begin{align}
& \begin{pmatrix} Z \\ Z' \end{pmatrix} = \begin{pmatrix} \cos \chi & \sin \chi \\ - \sin \chi & \cos \chi \end{pmatrix} \begin{pmatrix} \tilde{Z} \\ \tilde{Z}' \end{pmatrix}, \\
& \tan 2 \chi =  \frac{2 \Delta M^2 }{m^2_{Z_{\rm SM}} - \tilde{m}^2_{Z'}}.
\end{align}
Mass eigenvalues are also obtained as 
\begin{equation}
m^2_{Z, Z'} = \frac{1}{2} (m^2_{Z_{\rm SM}}+ \tilde{m}_{Z'}^2) \pm \frac{1}{2} \sqrt{(m^2_{Z_{\rm SM}} - \tilde{m}_{Z'}^2)^2 + 4 \Delta M^4}.
\end{equation}
{Here we briefly discuss constraints on $Z$-$Z'$ mixing from electroweak precision test (EWPT). 
The mixing is typically constrained as $|\sin \chi| \lesssim 10^{-3}-10^{-4}$ when $Z'$ boson comes from local $U(1)$ symmetry 
as subgroup of GUT symmetry like $SO(10)$ and $E_6$~\cite{Erler:2009jh}.
EWPT constraint is discussed for $U(1)_{L_\mu - L_\tau}$ case in ref.~\cite{Heeck:2011wj} where we find $-0.0008 \lesssim g_X \sin \chi \lesssim 0.0003$.
We apply the constraint as a reference under $m_{Z}^2 \ll m_{Z'}^2$ mass hierarchy for neutral gauge bosons. 
In our case the constraint approximately becomes
\begin{align}
\frac{g_X^2 m_Z v \cos \beta}{m_{Z'}^2} \lesssim 8 \times 10^{-4}.
\end{align}
In Fig.~\ref{fig:ZZpMixing}, we show the region excluded by EWPT for $U(1)_{L_\mu - L_\tau}$ with $\tan \beta =5$.
It is shown that $\mathcal{O}(1)$ gauge coupling is allowed if $Z'$ mass satisfies 1700 GeV $\lesssim m_{Z'}$ where the allowed region is also safe from other experimental constraints.
We expect similar $Z$-$Z'$ mixing constraint in the other cases, $U(1)_{L_e-L_\mu}$ and $U(1)_{L_e-L_\tau}$. 
This constraint restricts $v_\varphi \simeq m_{Z'}/g_X$ and can affect Majorana mass of $\nu_R$. 
However it does not provide much impact in fitting neutrino mass since we also have Yukawa couplings that are free parameters.
In our analysis below we consider heavy $Z'$ and assume $Z$-$Z'$ mixing is negligibly small to avoid EWPT.
}

\subsection{Charged lepton mass}

Mass matrix of charged lepton is generated via Yukawa interactions in Eq.~\eqref{eq:Lagrangian} after electroweak symmetry breaking.
Thus the mass term is given by
\begin{equation}
\overline{ \ell_L} (M_e)_{\ell \ell'} \ell'_R + h.c. =  \overline{ \ell_L} \left(  \frac{v_1}{\sqrt{2}} y_{\ell_1} + \frac{v_2}{\sqrt{2}} y_{\ell_2} \right)_{\ell \ell'} \ell'_R + h.c. \, ,
\end{equation}
where $\ell(\ell') = \{e, \mu, \tau\}$. 
In general the mass matrix is not diagonal and its structure depends on the model.
The possible structures of the matrix $M_e$ are as follows
\begin{align}
& M^{\rm Model \, (1)}_e : \left( \begin{array}{ccc} \times & 0 & \times \\ 0 & \times & 0 \\ 0 & \times & \times \end{array} \right), \quad 
M^{\rm Model \, (2)}_e : \left( \begin{array}{ccc} \times & 0 & 0 \\ 0 & \times & \times \\ \times & 0 & \times \end{array} \right), \quad
M^{\rm Model \, (3)}_e : \left( \begin{array}{ccc} \times & \times & 0 \\ 0 & \times & \times \\ 0 & 0 & \times \end{array} \right), \nonumber \\ 
& M^{\rm Model \, (4)}_e : \left( \begin{array}{ccc} \times & 0 & 0 \\ \times & \times & 0 \\ 0 & \times & \times \end{array} \right), \quad
M^{\rm Model \, (5)}_e : \left( \begin{array}{ccc} \times & 0 & \times \\ \times & \times & 0 \\ 0 & 0 & \times \end{array} \right), \quad
M^{\rm Model \, (6)}_e : \left( \begin{array}{ccc} \times & \times & 0 \\ 0 & \times & 0 \\ \times & 0 & \times \end{array} \right), 
\label{eq:models}
\end{align} 
where $\times$ indicates a non-zero element.
The mass matrix can be diagonalized by bi-unitary matrices $V_{eL}$ and $V_{eR}$ transforming $\ell_L$ and $\ell_R$; { $M_d^{\rm diagonal} = V_{eL}^\dagger M_e V_{eR}$}.
Note that such transformation induces FCNC for interaction among $Z'$ and leptons due to lepton flavor dependent charges and it is strongly constrained by LFV search experiments~\cite{Asai:2018ocx}.
{
Here we illustrate a LFV constraint in a simple case. 
Firstly $Z'$ interaction for charged leptons is written by
\begin{equation}
\mathcal{L}_{\rm Z'\ell \bar \ell}  = g_X Z'_\mu \left[ \overline{\ell_L} V^\dagger_{eL} Q_{L_\alpha - L_\beta} V_{eL} \gamma^\mu \ell_L + \overline{\ell_R} V^\dagger_{eR} Q_{L_\alpha - L_\beta} V_{eR} \gamma^\mu \ell_R \right],
\end{equation}
where $Q_{L_{\alpha -\beta}}$ indicates diagonal matrix whose elements are $L_\alpha - L_\beta$ charge, e.g. $Q_{L_e - L_\mu} = (1,-1,0)$ diagonal matrix.
The FCNC appears when $V^\dagger_{eL(eR)} Q_{L_\alpha - L_\beta} V_{eL(eR)}$ is not diagonal matrix.
As an example we consider $M_e$ of model (3) in Eq.~\eqref{eq:models} with $(M_e)_{23} =0$ for simplicity. 
Then, focusing on $e$-$\mu$ sector, the mass matrix is written by 
\begin{equation}
M_e^{e\mu} = \begin{pmatrix} M_e & \delta m \\ 0 & M_\mu \end{pmatrix}.
\end{equation}
Choosing $\delta m \ll M_\mu$ this matrix can be approximately diagonalized by 
\begin{align}
V^\dagger_{eL} M_e^{e \mu} V_{eR} \sim \begin{pmatrix} m_e & 0 \\ 0 & m_\mu \end{pmatrix}, \quad
 V_{eL} \sim \begin{pmatrix} 1 & \frac{\delta m}{M_\mu} \\ -\frac{\delta m}{M_\mu} & 1 \end{pmatrix}, \quad V_{eR} \sim \bf{1},
\end{align}
where $m_e \sim M_e$ and $m_\mu \sim M_\mu$.
Since model (3) is $U(1)_{L_e - L_\tau}$ case the LFV interaction is approximately given by
\begin{equation}
g_X \frac{\delta m}{m_\mu} Z'_\mu [  \bar{\mu} \gamma^\mu P_L e + \bar{e} \gamma^\mu P_L \mu].
\end{equation}
Then the experimental constraint for $\mu \to ee \bar e$ process is most stringent one for heavy $Z'$ case~\footnote{The case of light $Z'$ allowing $\ell_i \to \ell_j Z'$ decay is discussed in ref.~\cite{Asai:2018ocx} providing tighter constraints.}; that is induced via virtual $Z'$ as $\mu \to e Z'^* \to ee\bar{e}$.
We can estimate the decay width and branching ratio (BR) as
\begin{align}
& \Gamma_{\mu \to ee \bar e} \simeq \frac{g_X^4}{8 m_{Z'}^4} \left( \frac{\delta m}{m_\mu} \right)^2 \frac{m_\mu^5}{192 \pi^3}, \\ 
& BR(\mu \to ee \bar e) \simeq \frac{g_X^4}{g_2^2 G_F^2 m_{Z'}^4} \left( \frac{\delta m}{m_\mu} \right)^2,  
\end{align}
where $G_F$ is the Fermi constant. Applying the current limit $BR(\mu \to ee \bar e) \lesssim 10^{-12}$~\cite{SINDRUM:1987nra}, we obtain
\begin{equation}
g_X^4 \left( \frac{\delta m}{m_\mu} \right)^2 \left( \frac{\rm TeV}{m_{Z'}} \right)^4 \lesssim 6 \times 10^{-11}.
\end{equation}
Thus $\delta m/m_\mu$ should be very small as $\delta m/m_\mu \lesssim \mathcal{O}(10^{-5})$ when $g_X = \mathcal{O}(1)$ and $m_{Z'}$ is TeV scale. 
Note that we will obtain looser constraint when we consider LFV decay of $\tau$ lepton. 
In the numerical analysis below we  assume off-diagonal elements of the matrix is negligibly small and $V_{eL(eR)} \simeq 1$ for simplicity.
}

\subsection{Neutrino mass}

After spontaneous symmetry breaking we obtain Dirac mass $M_D$ between $\nu_L$ and $\nu_R$ ($\overline{\nu_L} M_D \nu_R$) as well as Majorana mass $M_R$ of $\nu_R$.
These mass matrices are written by
\begin{align}
M_D & =  \frac{y_{\nu_1} v_1}{\sqrt{2}} + \frac{y_{\nu_2} v_2}{\sqrt{2}}, \\
M_R & = M_0 + \frac{y_M v_\varphi}{\sqrt{2}} + \frac{\tilde{y}_M v_\varphi}{\sqrt{2}}.
\end{align}
Structure of these matrices is different for each model. 
For $M_D$ we find structure as follows:
\begin{align}
& M^{\rm Model \, (1)}_D : \left( \begin{array}{ccc} \times & 0 & 0 \\ 0 & \times & \times \\ \times & 0 & \times \end{array} \right), \quad 
M^{\rm Model \, (2)}_D : \left( \begin{array}{ccc} \times & 0 & \times \\ 0 & \times & 0 \\ 0 & \times & \times \end{array} \right), \quad
M^{\rm Model \, (3)}_D : \left( \begin{array}{ccc} \times & 0 & 0 \\ \times & \times & 0 \\ 0 & \times & \times \end{array} \right), \nonumber \\
& M^{\rm Model \, (4)}_D : \left( \begin{array}{ccc} \times & \times & 0 \\ 0 & \times & \times \\ 0 & 0 & \times \end{array} \right), \quad
M^{\rm Model \, (5)}_D : \left( \begin{array}{ccc} \times & \times & 0 \\ 0 & \times & 0 \\ \times & 0 & \times \end{array} \right), \quad
M^{\rm Model \, (6)}_D : \left( \begin{array}{ccc} \times & 0 & \times \\ \times & \times & 0 \\ 0 & 0 & \times \end{array} \right).
\end{align} 
We also find structure of $M_R$ for each model such that
\begin{align}
\label{eq:MR-structure}
& M^{\rm Model \, (1,2)}_R : \left( \begin{array}{ccc} 0 & \times & \times \\ \times & 0 & \times \\ \times & \times & \times \end{array} \right), \quad 
 M^{\rm Model \, (3,4)}_R : \left( \begin{array}{ccc} 0 & \times & \times \\ \times & \times & \times \\ \times & \times & 0 \end{array} \right), \quad 
 M^{\rm Model \, (5,6)}_R : \left( \begin{array}{ccc} \times & \times & \times \\ \times & 0 & \times \\ \times & \times & 0 \end{array} \right),  
\end{align}
where the structure is the same when a gauge symmetry of models is identical.

The neutrino mass in type-I seesaw model is given by
\begin{align}
m_\nu = - M_D M_R^{-1} M_D^T.
\end{align}
We then write active neutrino mass matrix by
\begin{align}
m_\nu  = \kappa \tilde{m}_\nu,
\label{eq:mnu-redef}
\end{align}
where $\kappa$ has mass dimension and $\tilde m_\nu$ is dimensionless.
The neutrino mass matrix $m_\nu$ is diagonalized by a unitary matrix $V_{\nu}$ by $D_\nu=|\kappa| \tilde D_\nu= V_{\nu}^T m_\nu V_{\nu}=|\kappa| V_{\nu}^T \tilde m_\nu V_{\nu}$.
Then $|\kappa|$ is determined by
\begin{align}
(\mathrm{NO}):\  |\kappa|^2= \frac{|\Delta m_{\rm atm}^2|}{\tilde D_{\nu_3}^2-\tilde D_{\nu_1}^2},
\quad
(\mathrm{IO}):\  |\kappa|^2= \frac{|\Delta m_{\rm atm}^2|}{\tilde D_{\nu_2}^2-\tilde D_{\nu_3}^2},
\label{eq:kappa}
 \end{align}
where $\Delta m_{\rm atm}^2$ is atmospheric neutrino mass-squared splitting, and NO and IO respectively represent the normal ordering and the inverted ordering of neutrino mass eigenvalues. 
Subsequently, the solar mass squared splitting can be written in terms of $|\kappa|$ as follows:
\begin{align}
\Delta m_{\rm sol}^2=  |\kappa|^2 ({\tilde D_{\nu_2}^2-\tilde D_{\nu_1}^2}),
 \end{align}
 which can be compared to the observed value.
 %
The observed mixing matrix is defined by $U=V^\dag_{eL} V_\nu$~\footnote{We can approximate $V_{eL} \simeq 1$ to suppress LFV as discussed in previous subsection. }, where
it is parametrized by three mixing angles $\theta_{ij} (i,j=1,2,3; i < j)$, one CP violating Dirac phase $\delta_{CP}$,
and two Majorana phases $\alpha_{21},\alpha_{31}$ as follows:
\begin{equation}
U = 
\begin{pmatrix} c_{12} c_{13} & s_{12} c_{13} & s_{13} e^{-i \delta_{CP}} \\ 
-s_{12} c_{23} - c_{12} s_{23} s_{13} e^{i \delta_{CP}} & c_{12} c_{23} - s_{12} s_{23} s_{13} e^{i \delta_{CP}} & s_{23} c_{13} \\
s_{12} s_{23} - c_{12} c_{23} s_{13} e^{i \delta_{CP}} & -c_{12} s_{23} - s_{12} c_{23} s_{13} e^{i \delta_{CP}} & c_{23} c_{13} 
\end{pmatrix}
\begin{pmatrix} 1 & 0 & 0 \\ 0 & e^{i \frac{\alpha_{21}}{2}} & 0 \\ 0 & 0 &  e^{i \frac{\alpha_{31}}{2}} \end{pmatrix}.
\end{equation}
Here, $c_{ij}$ and $s_{ij}$ stand for $\cos \theta_{ij}$ and $\sin \theta_{ij}$ ($i,j=1-3$), respectively. 
The matrix $U$ is identified as the PMNS matrix~\cite{Pontecorvo:1957qd,Maki:1962mu}.
Then, each of the mixings is given in terms of the component of $U$ as follows:
\begin{align}
\sin^2\theta_{13}=|U_{e3}|^2,\quad 
\sin^2\theta_{23}=\frac{|U_{\mu3}|^2}{1-|U_{e3}|^2},\quad 
\sin^2\theta_{12}=\frac{|U_{e2}|^2}{1-|U_{e3}|^2}.
\end{align}
The Dirac phase  $\delta_{CP}$ is given by computing  the Jarlskog invariant as follows:
\begin{align}
\sin \delta_{CP} &= \frac{\text{Im} [U_{e1} U_{\mu 2} U_{e 2}^* U_{\mu 1}^*] }{s_{23} c_{23} s_{12} c_{12} s_{13} c^2_{13}} ,\quad
\cos \delta_{CP} = -\frac{|U_{\tau1}|^2 -s^2_{12}s^2_{23}-c^2_{12}c^2_{23}s^2_{13}}{2 c_{12} s_{12} c_{23} s_{23}s_{13}} ,
\end{align}
where $\delta_{CP}$ be subtracted from $\pi$ if $\cos \delta_{CP}$ is negative.
Majorana phase $\alpha_{21},\ \alpha_{31}$ are found as
\begin{align}
&
\sin \left( \frac{\alpha_{21}}{2} \right) = \frac{\text{Im}[U^*_{e1} U_{e2}] }{ c_{12} s_{12} c_{13}^2} ,\quad
  \cos \left( \frac{\alpha_{21}}{2} \right)= \frac{\text{Re}[U^*_{e1} U_{e2}] }{ c_{12} s_{12} c_{13}^2}, \
%
,\\
&
 \sin \left(\frac{\alpha_{31}}{2}  - \delta_{CP} \right)=\frac{\text{Im}[U^*_{e1} U_{e3}] }{c_{12} s_{13} c_{13}},
\quad 
 \cos \left(\frac{\alpha_{31}}{2}  - \delta_{CP} \right)=\frac{\text{Re}[U^*_{e1} U_{e3}] }{c_{12} s_{13} c_{13}},
\end{align}
where $\alpha_{21}/2,\ \alpha_{31}/2-\delta_{CP}$
are subtracted from $\pi$, when $ \cos \left( \frac{\alpha_{21}}{2} \right),\  \cos \left(\frac{\alpha_{31}}{2}  - \delta_{CP} \right)$ are negative.
In addition, the effective mass for the neutrinoless double beta decay is given by
\begin{align}
\langle m_{ee}\rangle=|\kappa||\tilde D_{\nu_1} \cos^2\theta_{12} \cos^2\theta_{13}+\tilde D_{\nu_2} \sin^2\theta_{12} \cos^2\theta_{13}e^{i\alpha_{21}}+\tilde D_{\nu_3} \sin^2\theta_{13}e^{-2i\delta_{CP}}|,
\end{align}
where its observed value could be tested in experiments such as KamLAND-Zen~\cite{KamLAND-Zen:2022tow}, LEGEND~\cite{LEGEND:2017cdu} and nEXO~\cite{nEXO:2017nam}. 

\section{Numerical analysis}

 \begin{figure}[tb]
 \begin{center}
\includegraphics[width=6cm]{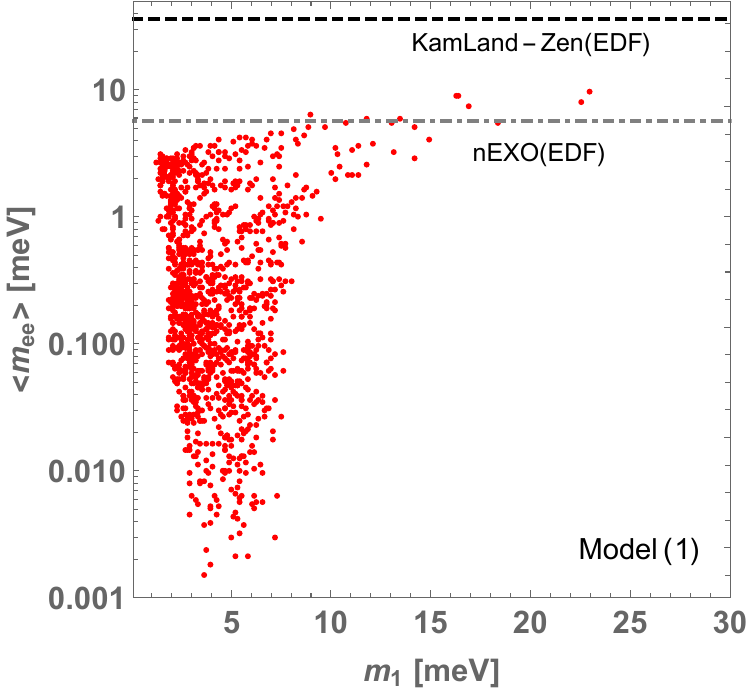}  \
\includegraphics[width=6cm]{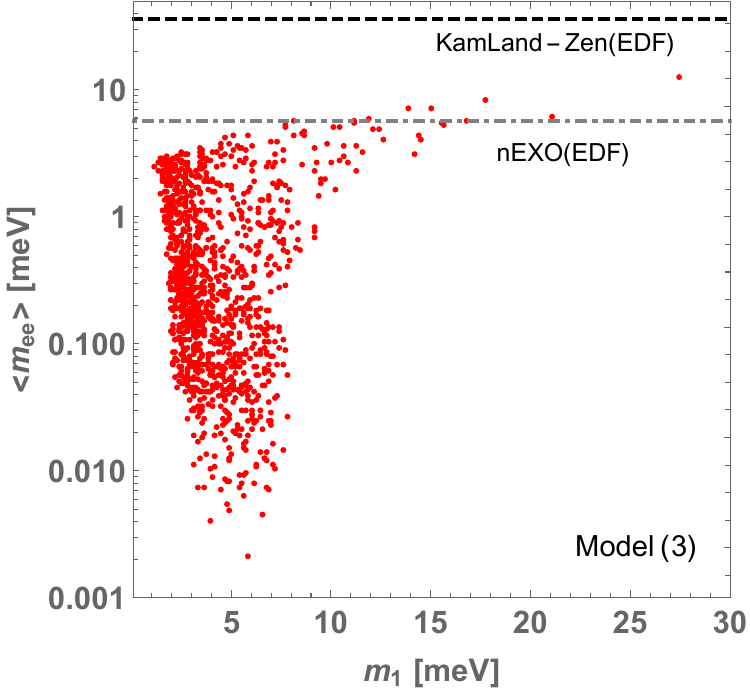} \\ \vspace{5mm}
\includegraphics[width=6cm]{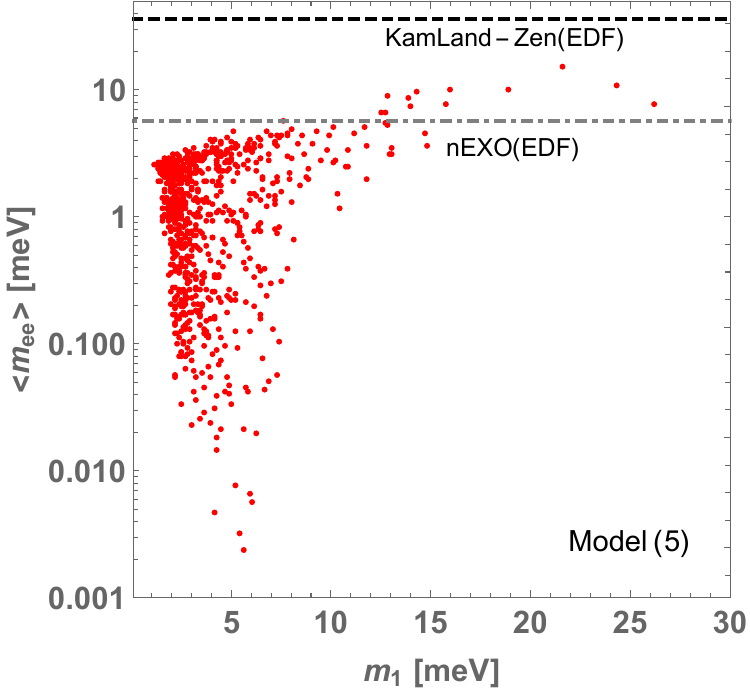} \
 \includegraphics[width=6cm]{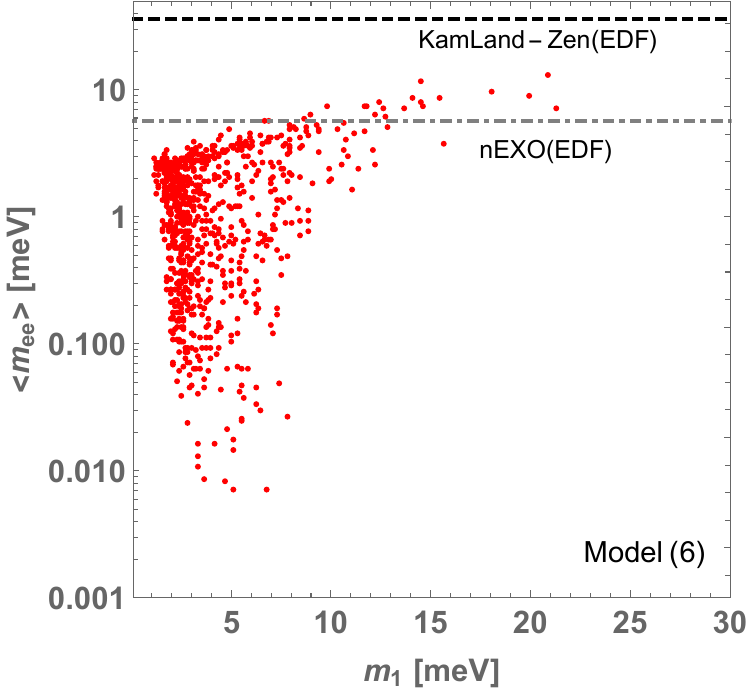}  
 \caption{The predicted values on $\{m_1, \langle m_{ee} \rangle \}$ plane for allowed parameter points in each model.}
\label{fig:m1-mee}
\end{center}
\end{figure}

 \begin{figure}[tb]
 \begin{center}
\includegraphics[width=6cm]{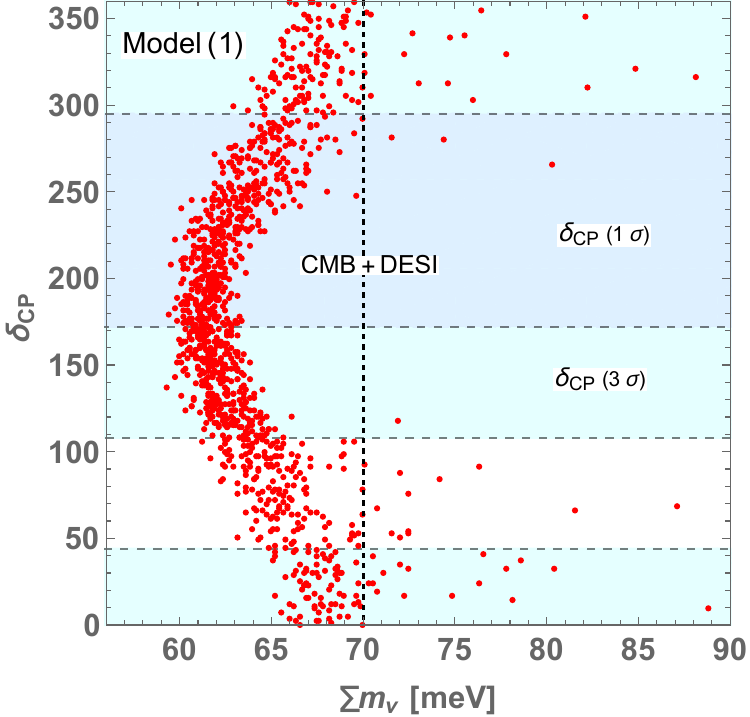}  \
\includegraphics[width=6cm]{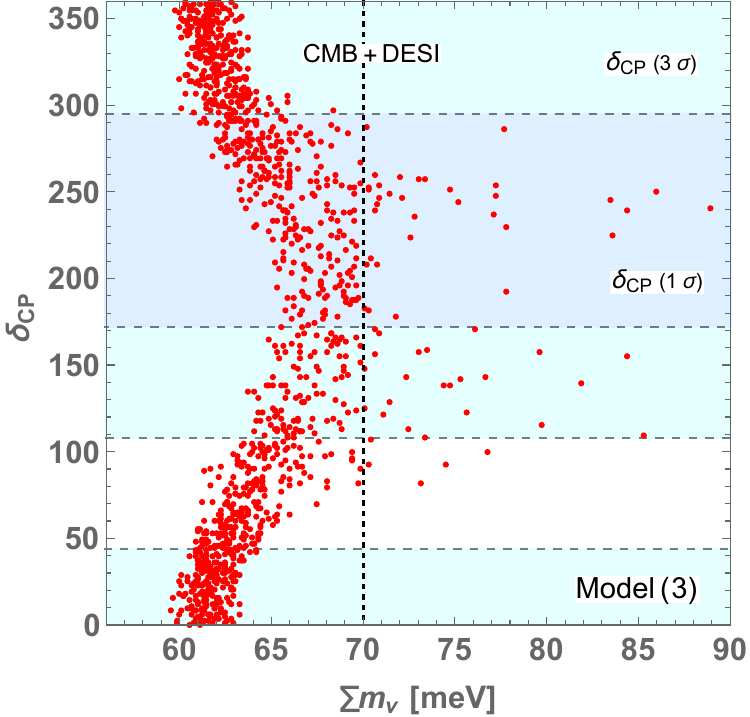} \\ \vspace{5mm}
\includegraphics[width=6cm]{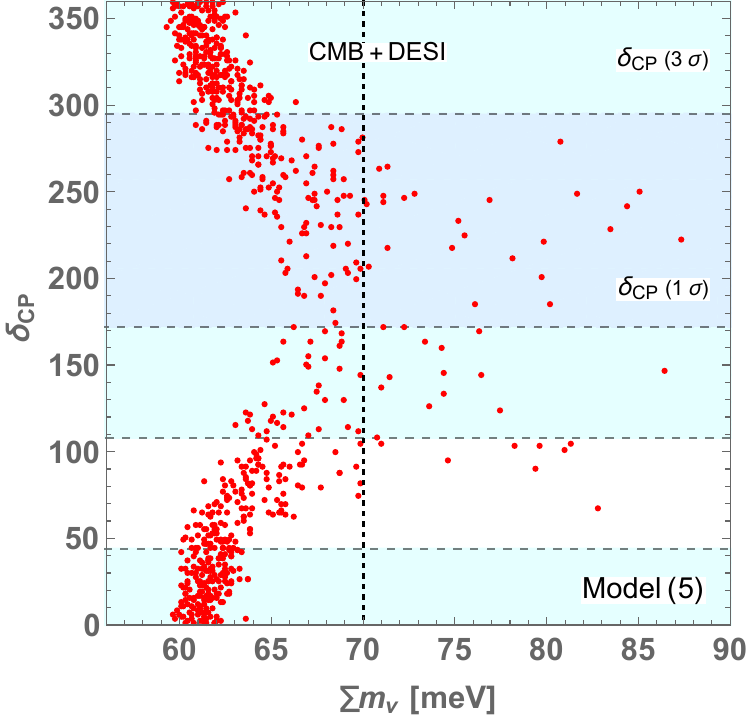} \
 \includegraphics[width=6cm]{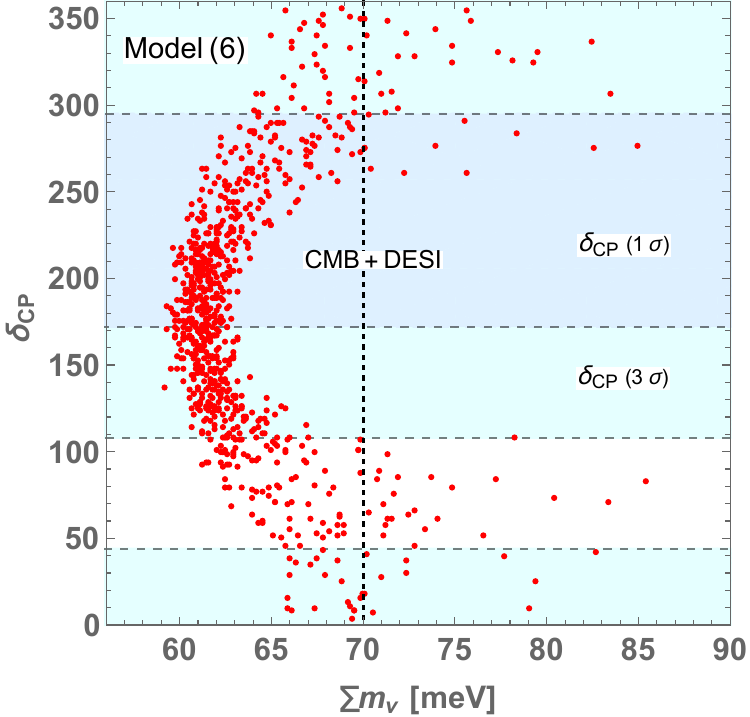}  
 \caption{The predicted values on $\{\sum m_\nu, \delta_{\rm CP} \}$ plane for allowed parameter points in each model. 
 The light-blue(cyan) region indicates 1(3)$\sigma$ range of $\delta_{\rm CP}$ value.}
\label{fig:sum-delCP}
\end{center}
\end{figure}

 \begin{figure}[tb]
 \begin{center}
\includegraphics[width=6cm]{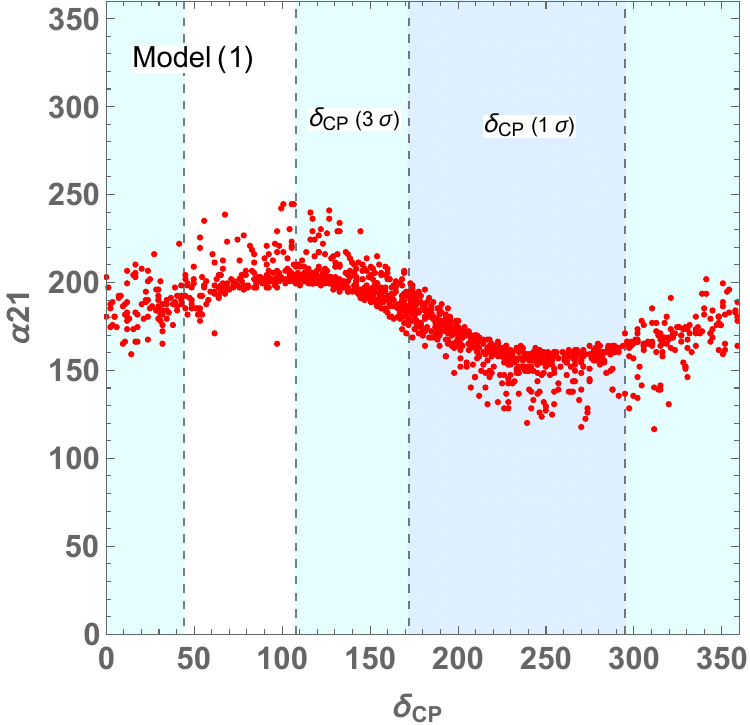}  \
\includegraphics[width=6cm]{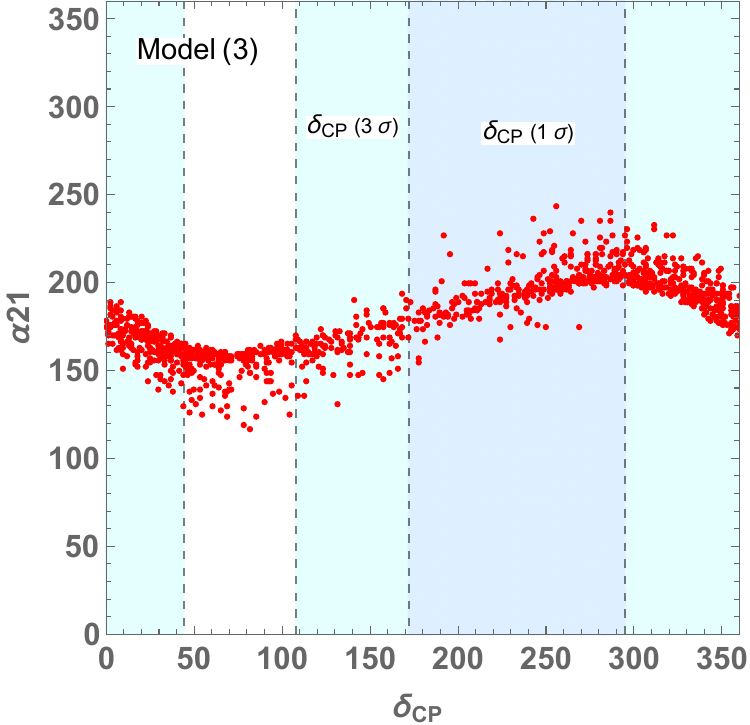} \\ \vspace{5mm}
\includegraphics[width=6cm]{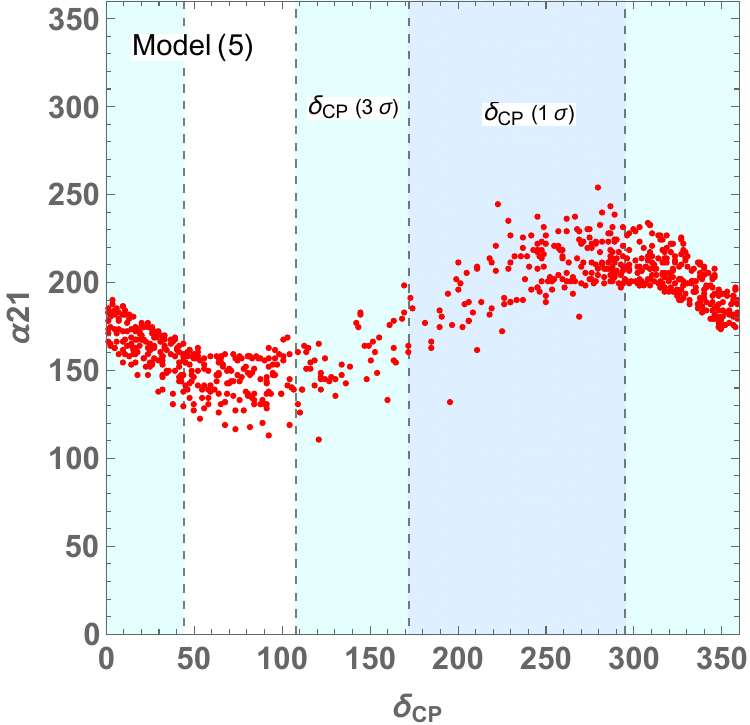} \
 \includegraphics[width=6cm]{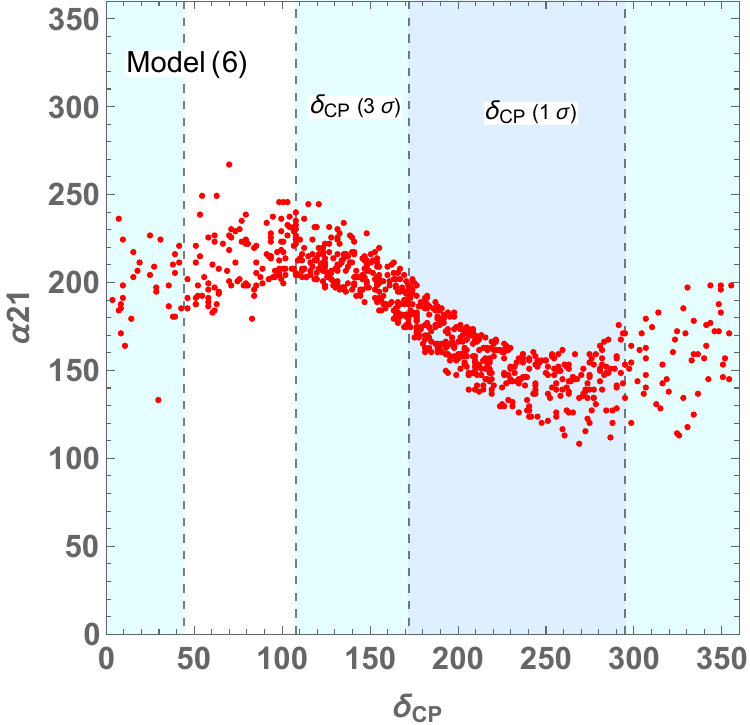}  
 \caption{The predicted values on $\{\delta_{\rm CP}, \alpha_{21} \}$ plane for allowed parameter points in each model. 
 The light-blue(cyan) region also indicates 1(3)$\sigma$ range of $\delta_{\rm CP}$ value.}
\label{fig:delCP-a21}
\end{center}
\end{figure}

In this section we discuss neutrino observables such as masses, mixings and CP-phases in our models.
These values are numerically estimated since structures of the neutrino mass matrix are not very simple to get analytic solutions for neutrino observables.  
In numerical calculation we modify neutrino mass matrix as 
\begin{equation}
m_\nu = - \frac{(M_D)_{11}^2 }{(M_R)_{kk}} \tilde{M}_D \tilde{M}_R^{-1} \tilde{M}_D^T,
\end{equation}
where $\tilde{M}_D \equiv M_D/(M_D)_{11}$ and $\tilde{M}_R= M_R/(M_R)_{kk}$ with $(M_R)_{kk}$ being a non-zero diagonal element of $M_R$ in Eq.~\eqref{eq:MR-structure}.
The scaling factor $\frac{(M_D)_{11}^2 }{(M_R)_{kk}}$ is identified as $\kappa$ in Eq.~\eqref{eq:mnu-redef} that will be evaluated by Eq.~\eqref{eq:kappa}.
Then we scan the element of $\tilde{M}_D$ and $\tilde{M}_R$ in the range of
\begin{equation}
\left| (\tilde{M}_{D,R})_{ij} \right| \in [0.1, 10],  \quad \left( (\tilde{M}_D)_{11} =1, \quad (\tilde{M}_R)_{kk} =1, \right)
\end{equation}
where we consider these matrix elements do not have large hierarchy.
Also note that the elements are complex in general and we remove phases of diagonal elements of matrix $M_D$ by redefining phases of lepton fields without loss of generality.
The neutrino observables are numerically estimated using the formulas in the previous section to explore the tendency in each model.
In the analysis we adopt NuFit 5.2 neutrino data~\cite{Esteban:2020cvm} for the values of $\{\sin \theta_{12}, \sin \theta_{23}, \sin \theta_{13}, \Delta m^2_{\rm atm}, \Delta m^2_{\rm sol}\}$ within 3$\sigma$ range as
\begin{align}
& 0.270 < \sin^2 \theta_{12} < 0.341, \quad  0.406 < \sin^2 \theta_{23} < 0.620, \quad  0.02029 < \sin^2 \theta_{13} < 0.02391, \nonumber \\
& 2.428 \times 10^{-3} \ {\rm eV}^2 < |\Delta m_{\rm sol}^2| < 2.597 \times 10^{-3} \ {\rm eV}^2, \nonumber \\
&  6.82 \times 10^{-5} \ {\rm eV}^2 < \Delta m^2_{\rm sol} < 8.03 \times 10^{-5} \ {\rm eV}^2.
\end{align}
Then we search for values of $(\tilde{M}_{D,R})_{ij}$ reproducing these 3$\sigma$ ranges of the observables. 
Here we focus on NO case and check if we can get sum of neutrino masses $\sum m_\nu$ satisfying constraint of CMB+DESI, $\sum m_\nu < 70$ meV.
As a result, we find that model (2) and (4) are disfavored to fit the neutrino data within 3$\sigma$ while the other models have allowed parameter sets to fit the data.
It is also found that we don't have clear correlation between neutrino mixing angles and the other observables due to increased free parameters compared to minimal scenarios.
For the allowed models we show some observables such as sum of masses and phases as our predictions.

In Fig.~\ref{fig:m1-mee}, we show the $\langle m_{ee} \rangle$ and $m_1$ (the lightest neutrino mass) for allowed parameter points in each model.
The horizontal lines shows current constraint from KamLand-Zen~\cite{KamLAND-Zen:2022tow} and future prospect in nEXO~\cite{nEXO:2017nam} with energy-density functional (EDF) theory for nuclear matrix element.
The distributions are almost similar in these models and some points can be tested in future neutrinoless double beta decay experiments.

 \begin{figure}[tb]
 \begin{center}
\includegraphics[width=6cm]{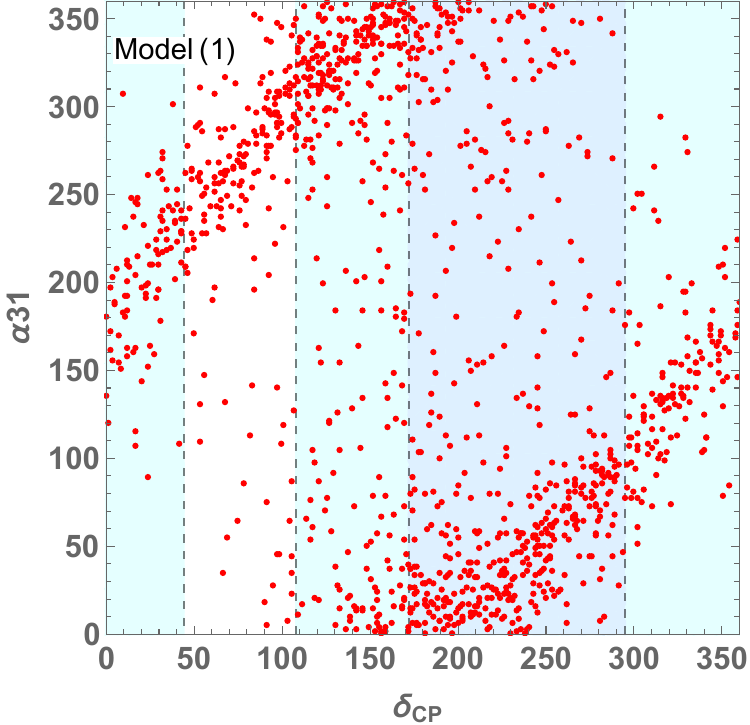}  \
\includegraphics[width=6cm]{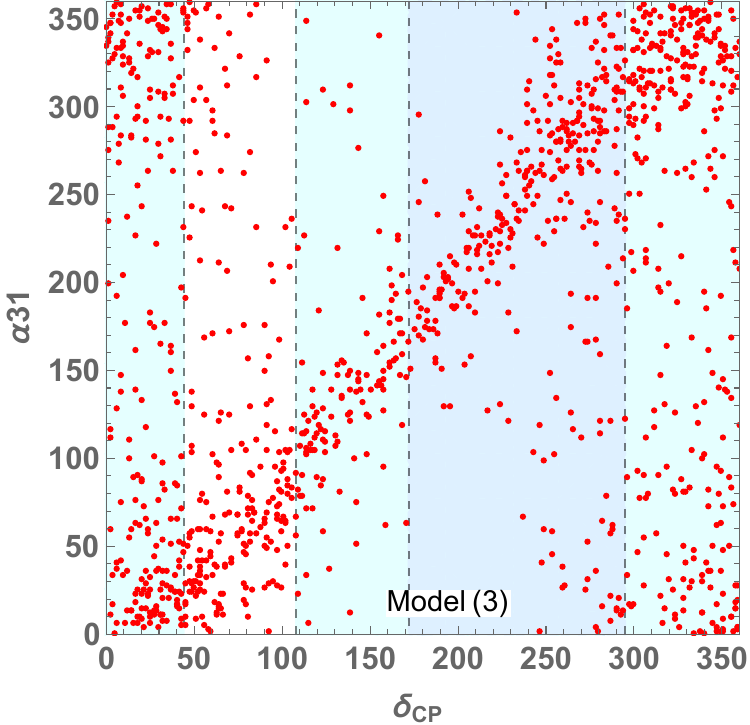} \\ \vspace{5mm}
\includegraphics[width=6cm]{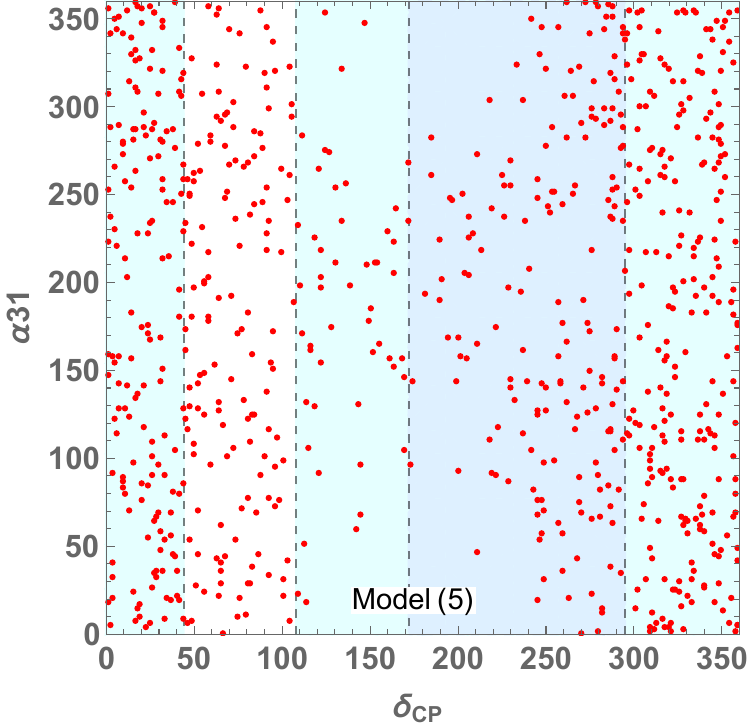} \
 \includegraphics[width=6cm]{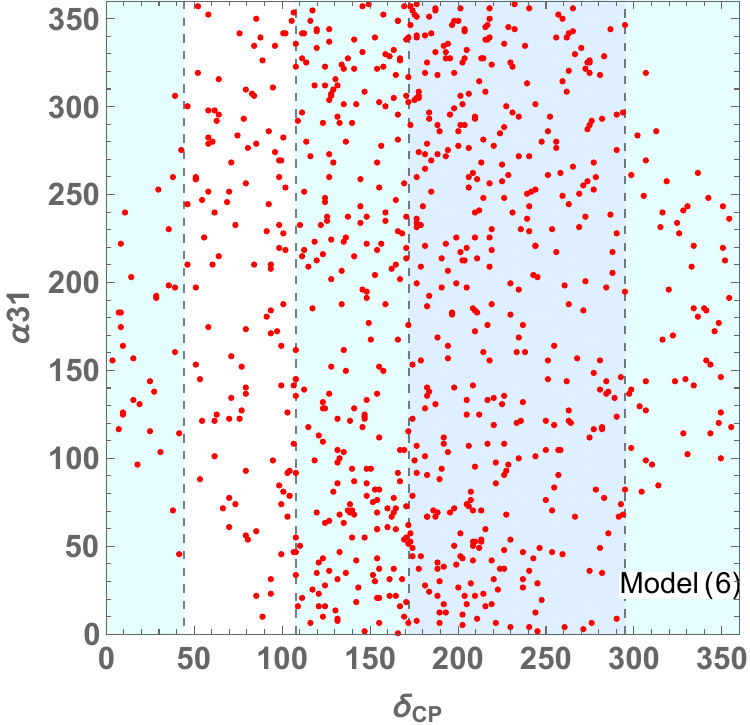}  
 \caption{The predicted values on $\{\delta_{\rm CP}, \alpha_{31} \}$ plane for allowed parameter points in each model. 
 The light-blue(cyan) region is the same as Fig.~\ref{fig:delCP-a21}.}
\label{fig:delCP-a31}
\end{center}
\end{figure}

In fig.~\ref{fig:sum-delCP}, we show $\sum m_\nu$ and $\delta_{\rm CP}$ for allowed parameter points in each model.
The vertical line indicates the upper limit by CMB+DESI data.
We find some correlation between $\sum m_\nu$ and $\delta_{\rm CP}$ where models (1) and (6) [(3) and (5)] provide similar behavior.
As we see, they show ditinguishable behavior where one tends to provide small $\sum m_\nu$ around $\delta_{\rm CP} \sim 200^{\circ}$ and 
the other one tends to give small $\sum m_\nu$ around $\delta_{\rm CP} \sim 0^{\circ}(360^{\circ})$.
We also indicate 1(3)$\sigma$ range of $\delta_{\rm CP}$ value by the light-blue(cyan) region.
These two types of distributions could be distinguished if we have more precision for $\sum m_\nu$ and $\delta_{\rm CP}$ in future.

\begin{figure}[tb]
 \begin{center}
\includegraphics[width=6cm]{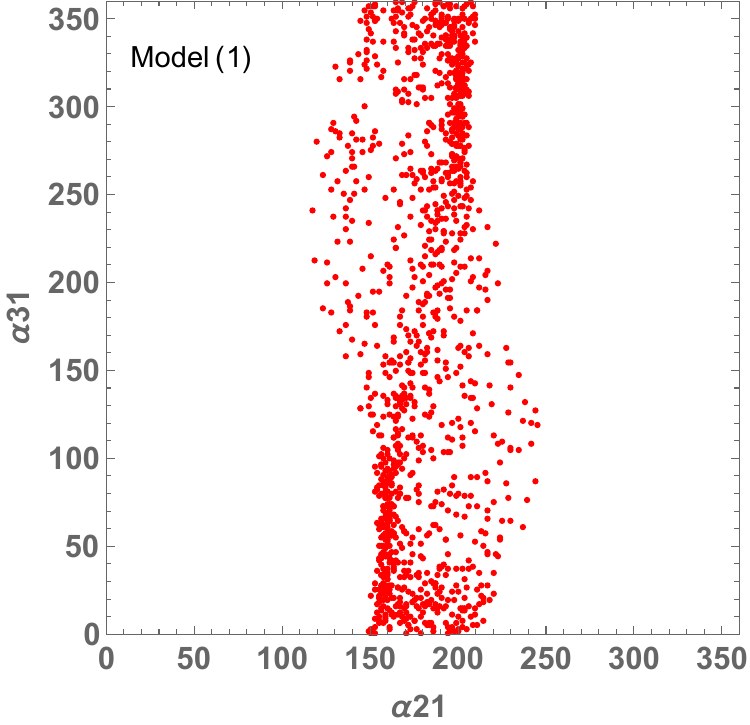}  \
\includegraphics[width=6cm]{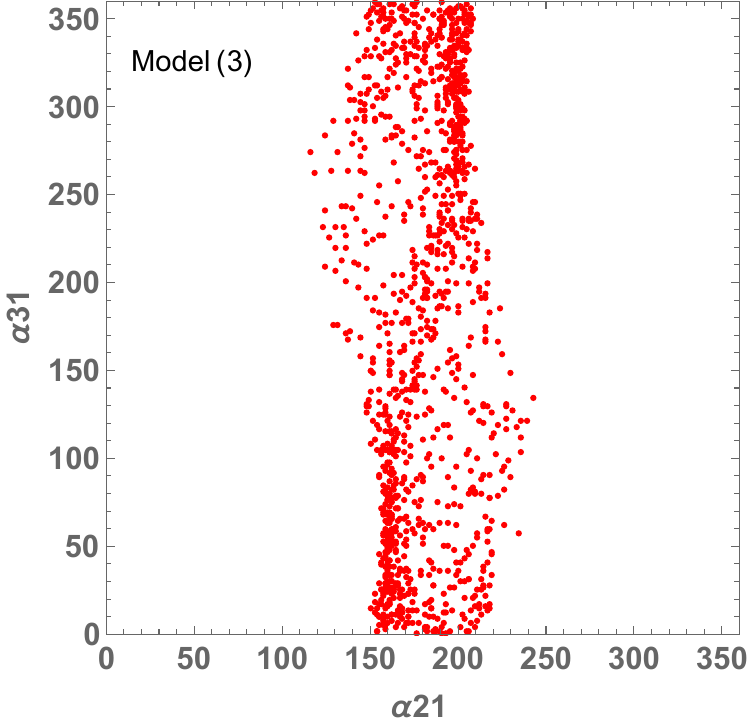} \\ \vspace{5mm}
\includegraphics[width=6cm]{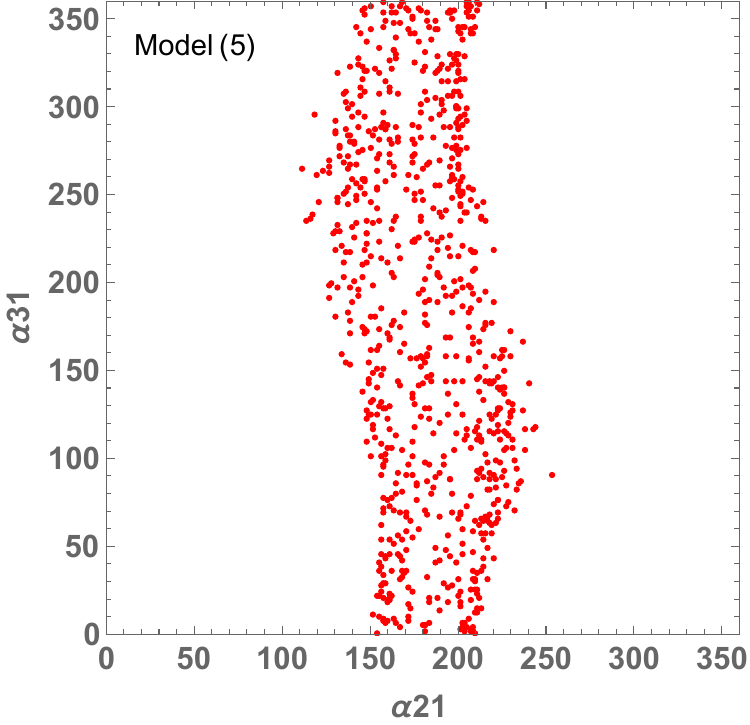} \
 \includegraphics[width=6cm]{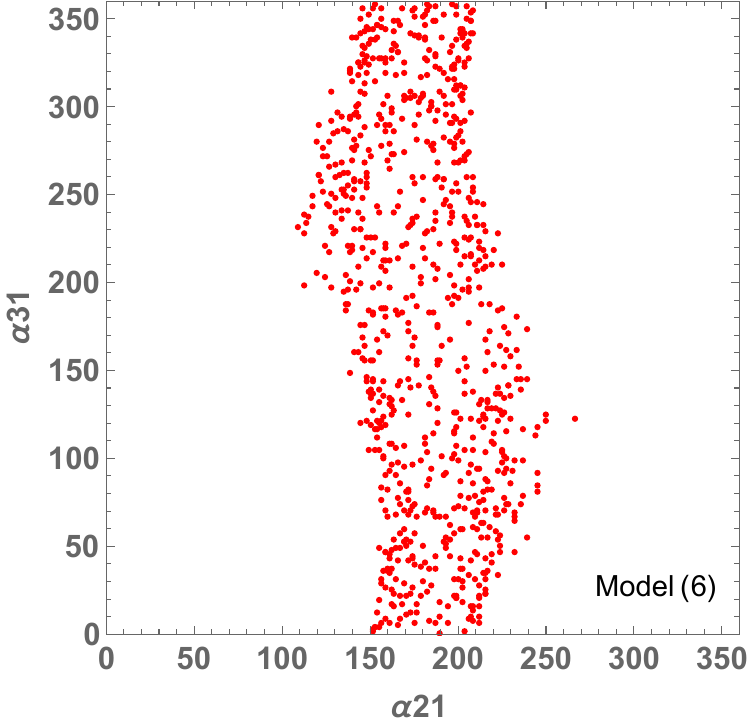}  
 \caption{The predicted values on $\{\alpha_{21}, \alpha_{31} \}$ plane for allowed parameter points in each model.}
\label{fig:a21-a31}
\end{center}
\end{figure}

Fig.~\ref{fig:delCP-a21} and \ref{fig:delCP-a31} show $\delta_{\rm CP}$-$\alpha_{21}$ and  $\delta_{\rm CP}$-$\alpha_{31}$ values for allowed parameter points in each model.
We find some correlations among phases where correlations of $\delta_{\rm CP}$-$\alpha_{21}$ in models (1) and (6) [(3) and (5)] provide similar behavior.
In each case the value of $\alpha_{21}$ is concentrated around 100-250 degrees.
On the other hand correlations of  $\delta_{\rm CP}$-$\alpha_{31}$ are different in each model. 
In these plots we show 1(3)$\sigma$ range of $\delta_{\rm CP}$ value by the light-blue(cyan) region as in the Fig.~\ref{fig:sum-delCP}.
In addition, we also show correlation between $\alpha_{21}$ and $\alpha_{31}$ in Fig.~\ref{fig:a21-a31} where models (1) and (6) [(3) and (5)] provide similar behavior.

\section{Summary and discussion}

We have discussed neutrino observables in $U(1)_{L_\alpha - L_\beta}$ models where we introduced second Higgs doublet and one singlet scalar fields with non-zero charges under new $U(1)$, as extensions of minimal models that have only one new scalar field.
The six models are characterized by the choice of $U(1)_{L_\alpha - L_\beta}$ gauge symmetry and its charge assignment to the second Higgs doublet.
We have then formulated scalar sector, neutral gauge bosons, charged lepton mass matrix and neutrino mass matrix in the models.
The structure of neutrino mass matrix is still restricted by the gauge symmetry but number of free parameters is increased compared to minimal cases.

In this work we have focused on the neutrino sector and numerically analyzed neutrino observables.
Then we have found four models can fit the current neutrino data in 3$\sigma$ and satisfy constraint on sum of neutrino mass including recent DESI data and Planck data
for normal ordering that excludes minimal models under $\Lambda$CDM cosmological model.
Although we don't have correlation between neutrino mixing angle and other observable due to increased free parameters 
we have found some correlations between some observables like sum of neutrino mass and CP phases.
In particular we have found two types of distinguishable distributions on $\{\sum m_\nu, \delta_{\rm CP} \}$ plane that could be tested in future increasing precision for these observables.
There are also some correlations among CP phases although it is difficult to measure them directly.
For further distinguishing models we would need to explore Higgs and $Z'$ physics, for example, searching for collider signals which are specific in models.
Analysis of these physics is beyond the scope of this paper and it is left for future works.

\section*{Acknowledgments}
The work was also supported by the Fundamental Research Funds for the Central Universities (T.~N.). 

\end{document}